# Mechanically Regulated Cranial Growth in Infancy: A Computational Approach to Predicting Craniosynostosis

Mahtab Vafaeefar[1], Conall Quinn[1], Ted J. Vaughan[1]

[1] Biomechanics Research Centre (BMEC), School of Engineering, Institute for Health Discovery and Innovation, College of Science and Engineering, University of Galway, Ireland.

*Address for correspondence:*
Prof Ted J Vaughan
Professor in Biomedical Engineering,
Biomechanics Research Centre (BMEC)
Biomedical Engineering
University of Galway
Galway
Ireland
Phone: (+353) 91-493084
Email: ted.vaughan@universityofgalway.ie

Abstract

In early years of life, the cranium rapidly changes in size and shape to accommodate brain growth, primarily driven by mechanical stress from brain expansion. Developmental disorders such as premature fusion of sutures in craniosynostosis, disrupts normal growth process, leading to abnormal skull shapes. Thus, understanding the interplay between biomechanical forces, soft tissues, and individual bone plates is crucial for understanding their role in shaping infant skulls. This study develops a mechanically-driven growth model to simulate healthy cranial growth in the first year. The algorithm considers simultaneous and coupled growth of brain, cranial bones, sutures, with volumetric brain expansion as the primary driver, with strain-based feedback governing growth in bone and suture tissues. A bulk bone formation approach accounts for evolving mechanical properties, with elastic moduli of bone and sutures increasing monthly. The model was applied on individual fused sutures and skull dysmorphologies due to craniosynostosis were predicted, and results showed good agreement with clinically observations. Stress at bone-suture interfaces and elevated intracranial pressure under fused sutures highlighted biomechanical impacts due to the disorders. Sensitivity analysis explored how material properties and growth rates affect skull shape. This framework enhances understanding of cranial growth and supports treatment planning for craniosynostosis.

## 1 Introduction

The cranium is composed of multiple bone plates that are connected by sutures, which are made of fibrous connective tissue, allowing for movement during birth and early development. The peripheral suture tissue is crucial to new bone deposition sites during skull growth and development in infants [1]. Cranial and skull growth is a complex and highly regulated process that involves coordinated biological, chemical, and mechanical signalling to drive development of bones, sutures, and the underlying brain [2,3]. It is well-acknowledged that there are many genetic and epigenetic factors responsible for the skull growth. However, it is commonly recognised that the main mechanism that drives rapid growth of the cranium is large brain expansion during early years of life [1,4,5]. Sutural growth allows the skull to expand in response to brain growth by forming new bone at the edges of individual plates, while surface growth (or appositional growth) takes place at the surface of calvaria and remodels the overall bone shape and thickness [6]. While there is physical evidence demonstrating the influence of mechanical loads on the cranial growth [3], only a limited number of studies have systematically investigated the underlying mechano-regulatory growth mechanisms that contribute to cranial development in infants [2,6–8].

Craniosynostosis is a congenital condition characterized by the premature fusion of one or more cranial sutures, affecting approximately 1 in 2,000 live births. This early suture closure disrupts normal skull and brain growth, often resulting in abnormal head shapes, distinctive facial features [9–11], and even loss of vision or cognitive impairment due to the restricted skull expansion in areas where sutures have prematurely fused [12]. Typically, the skull expands in planes perpendicular to the sutures, but early fusion redirects growth parallel to the closed suture [11]. Correction techniques for craniosynostosis primarily aim to restore normal skull growth by reopening fused sutures and creating space for the expanding brain, restoring a more typical head shape and preventing further neurological complications. Treatments of this condition require multidisciplinary experts and involve invasive surgery in most cases. However, blood loss and surgical time are significant concerns for the young patients, and there is a need for less invasive yet efficient techniques [10,13]. Efficient treatment plans require a deep understanding of normal craniofacial development that relies on insight into how growth dynamics interplay within the cranium [14]. This highlights the importance of understanding the biomechanical contributions towards cranial growth, which includes the mechanical interactions between the brain growth, its forces on soft tissue and bone plates, that leads to the development of the final shape of the skull in infants [1].

Given the complexities of in vivo experimentation and indeed ethical challenges associated with studying infant cranial development, computational techniques such as finite element analysis (FEA) have become increasingly valuable for predicting cranial biomechanics and growth. At a cellular level, a computational model for mechanobiological bone formation mechanism in cranial vault has been developed, by coupling reaction-diffusion algorithms with structural mechanics [7]. However, this model did not consider volumetric growth of the domain, focusing solely on simulating bone formation at the cellular scale. At the organ level, many existing finite element models developed have not accounted for growth dynamics on the long-term biomechanical assessment [13,15–18], focusing instead on static evaluation of implants during pre-operative and immediate post-operative periods in craniosynostosis treatment simulations [16]. In one study, growth was simulated by applying either uniform or non-uniform surface pressure on the inner skull surface [4]. However, this model was not able to capture changes in skull shape that arise in dysmorphologies such as craniosynostosis, since the mechanical properties, size and shape of the skull was kept constant, with growth dynamics not being considered. In recent studies, physical growth phenomena have been incorporated [2,6,9]. Some of these approaches have included volumetric growth of the brain being considered through a thermal expansion on the intercranial volume (ICV), in which the intercranial vault pressure was studied with the brain expansion [2,8,12,19,20]. In other approaches, growth has been modelled as gradual bone formation process at an element level at specific radius from adjacent bone [8,20], or bulk bone formation within suture tissue, where the bulk elastic modulus of the of the suture/craniotomy was increased [2], which was computationally less expensive. Both approaches predicted the overall morphology of the skull after growth, with differences in predicted levels of contact pressure on brain [12]. The bone formation modelling approach for growth was later extended to human skulls to evaluate the efficacy of different correction techniques for sagittal craniosynostosis [9,21]. Using this algorithm, sensitivity analysis on post-operative calvarial growth in sagittal craniosynostosis revealed that the most impactful parameter on the predicted skull morphology was the elastic modulus of the craniotomies [12]. However, these models only considered short term growth periods [20], and brain growth was implemented as multiple volume increase intervals, not as a gradual continuous volumetric growth process, as it inherently is [2,21]. More advanced kinematic volumetric growth models, such as that developed by [6], have also been used to investigate the geometrical and dimensional constraints in skull growth. However, even in this approach, brain growth was modelled as an intracranial pressure applied uniformly to the inner skull surface, without considering changes in bone and suture mechanical properties during development. For effective and accurate predictions of cranial growth, robust, mathematically formulated kinematic growth models must be employed that appropriately incorporate continuous volumetric brain expansion, evolving mechanical properties of cranial tissues, and the complex biomechanical interactions between bone, sutures, and the underlying ICV.

The objective of this study is to develop a physically-based growth model to predict the skull and calvaria development within the first year of age, for healthy and craniosynostosis cases. The study presents a mechanically driven growth algorithm that simulates coupled, and gradual growth of skull and intercranial volume (ICV) or brain, where the skull growth was stimulated by the underlying brain tissue volumetric growth. Besides the volumetric brain growth, and mechanobiological skull growth, bulk bone formation was applied on the suture section, to account for the gradual bone formation on the bone-suture edges. Moreover, variation of the mechanical properties of bone over time, was considered in this study. Using the developed model, dimensional and morphological prediction of skull malformation in craniosynostosis, for individual suture fusion, was studied within the first 12 months of age. Sensitivity analysis was also performed to study the effect of model parameters on the outcome of simulations.

## 2 Methods

### 2.1 Model Geometry

Figure 1(a) shows the skull geometry that was approximated as an ellipsoid with cephalic index ($CI$) of 78, with length of 206 mm and width of 162 mm, created in Autodesk Inventor CAD software. The geometry consisted of 13 pieces of bone and the suture sections. Sutures were categorised as three sections, representing for directional growth of skull. Longitudinal and transverse sutures are responsible for longitudinal and transverse growth of skull, respectively, and the fontanelle sutures, that grow in both directions as shown in Figure 1(a). The directional growth implementation is discussed in more details in Section 2.3. As a measurement of the shape of skull cephalic index is clinically used [18], which is defined as the ratio of maximum width to the maximum length of the skull.

$$CI = \frac{maximal\ width}{maximal\ length} \cdot 100\% \qquad (1)$$

The model was discretised into reduced integration tetrahedral elements (C3D8R), with enhanced hourglass controls. A mesh convergence study was performed, by increasing element numbers and convergence was seen when to have been achieved once both normalised maximum displacement and a local stress had plateaued by $\pm 5\%$ alternations, with more information available in the supplementary document. The final model consisted of 42,032 of skull elements, and 31,200 of brain elements, resulting in the mesh density shown in Figure 1(b). The FE models were analysed using an implicit solver in ABAQUS (SIMULIA, Dassault Systémes), with the growth algorithm as a UMAT and material orientation as an ORIENT subroutine.

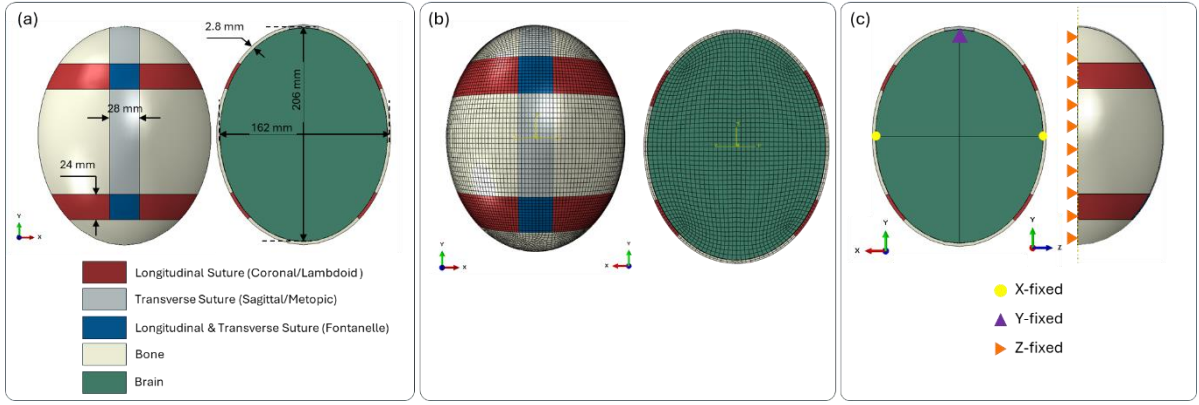

Figure 1 FE Model preparation; (a) Idealised skull and ICV geometry for FE model, (b) mesh intensity on material sections, (c) boundary conditions applied.

The boundary conditions and interactions on the FE model are shown in Figure 1(c). The ICV and skull surfaces are fixed from the bottom $xy$ plane, and to constrain in plane rotations, $X$-fixed and $Y$-fixed nodes on the ICV are fixed in constrained in $x$ and $y$ directions, respectively. The bone-suture interface was assumed to be perfectly connected. Contact interaction was established between the inner skull surface, including bone and sutures faces, with the ICV face. The contact was defined with normal contact stiffness of 50 N/mm, and a tangential friction coefficient of 0.1 with slip tolerance of 0.005 mm [21], with stabilisation coefficient of 0.01. These surfaced were initially in contact and were allowed movement during growth process.

### 2.2 Constitutive Models and Continuum Growth Theory

In addition to the theory of morphoelasticity [22], nonlinear field theories of mechanics have been used to mathematically simulate finite growth in elastic materials [6]. From continuum mechanics, the deformation matrix of the growing system $\boldsymbol{\varphi}$, maps the material point $X$ at time $t$ to the spatial point $x = \boldsymbol{\varphi}(X, t)$. According to finite strain theory in continuum mechanics, the deformation gradient $\boldsymbol{F}$ is described as the spatial gradient of the mapping function ($\boldsymbol{\varphi}$) that describes the motion of the continuum. Kinematically, finite growth was described through the decomposition of deformation gradient matrix, where it consists of an elastic deformation ($\boldsymbol{F}^e$), and a growth deformation ($\boldsymbol{F}^g$) term [6,23], as expressed in Equation (2).

$$\boldsymbol{F} = \boldsymbol{\nabla} \times \boldsymbol{\varphi} = \boldsymbol{F}^e . \boldsymbol{F}^g \tag{2}$$

Constitutively, the elastic contribution of each material was defined based on their strain energy density functions ($W$) [24,25]. Brain tissue was considered a Neo-Hookean hyperelastic material, while bone and suture sections were modelled as linear elastic. For a Neo-Hookean material, strain energy density is defined as,

$$W = \frac{\mu}{2}(I_1 - 3) - \mu \ln J + \frac{\lambda}{2}(\ln J)^2 \tag{3}$$

and for linear elastic material model is defined as,

$$W = \frac{1}{2}\lambda(tr(\boldsymbol{\varepsilon}))^2 - \mu\, tr(\boldsymbol{\varepsilon}^2) \tag{4}$$

in which $\mu$ and $\lambda$ are Lamé parameters, $I_1$ is the first invariant of the left Cauchy Green deformation tensor, $J$ is the Jacobian determinant, and $\boldsymbol{\varepsilon}$ is the infinitesimal (small) strain tensor. For a Neo-Hookean material, Cauchy stress can be calculated based on the left Cauchy-Green tensor $\boldsymbol{b}$ as,

$$\boldsymbol{\sigma} = \frac{\mu}{J}(\boldsymbol{b} - \boldsymbol{I}) - \frac{\lambda}{J}(\ln J)\boldsymbol{I} \tag{5}$$

For a linear elastic material, the Cauchy stress is a linear function of the small strain tensor, and can be calculated based on the elasticity tensor $\boldsymbol{C}$, known as Hooke's Law,

$$\boldsymbol{\sigma} = \boldsymbol{C}:\boldsymbol{\varepsilon} \tag{6}$$

*2.3    Anisotropic Growth Implementation*

Besides the elastic deformation contribution term, which is calculated based on finite strain theory, the growth term of the deformation gradient tensor was defined as,

$$\boldsymbol{F}^g = \begin{bmatrix} \boldsymbol{F}^g(1,1) & 0 & 0 \\ 0 & \boldsymbol{F}^g(2,2) & 0 \\ 0 & 0 & \boldsymbol{F}^g(3,3) \end{bmatrix} \tag{7}$$

The growth matrices in Equation (7) were defined individually in the principal directions of the local material orientations, for each respective tissue. Figure 2 shows model sections, and the local material orientations, represented as longitudinal, transverse and normal directions, for the bone and suture tissues. The growth matrix then can be written as:

$$\boldsymbol{F}^g = F^g_l\, \boldsymbol{l}\otimes\boldsymbol{l} + F^g_t\, \boldsymbol{t}\otimes\boldsymbol{t} + F^g_n\, \boldsymbol{n}\otimes\boldsymbol{n} \tag{8}$$

in which growth terms in local material orientation is updated in each increment as:

$$\boldsymbol{F}^g_{T+dT,i} = \Delta \boldsymbol{F}^g_{T,i} + \boldsymbol{F}^g_{T,i}, \quad i = l, t, n \tag{9}$$

where $\Delta \boldsymbol{F}^g_T$, in each material direction, is calculated based on the corresponding growth governing functions, at time $T$.

Suture growth along the edges involves the gradual development of new bone material at the margins of bone plates, progressing outward in a direction perpendicular to the edge [6]. According to this physiological framework, represented for the FE model in Figure 2, the metopic and sagittal sutures predominantly facilitated the skull's transverse growth (shown as grey patches), whereas the coronal and lambdoid sutures contributed to its elongation (indicated by red patches). The anterior and posterior fontanelles, located at the intersections of longitudinal and transverse sutures, are thought to support expansion in both key directions (shown as blue patches). To represent the material orientation on the skull part, a normal and two tangential directions on each element edge has been defined as orthotropic in-plane directions. In the bone and suture sections, a local material orientation was defined using an ORIENT subroutine. In the ICV section, the material orientation was the same as the global orientation, where linear growth was defined isotopically.

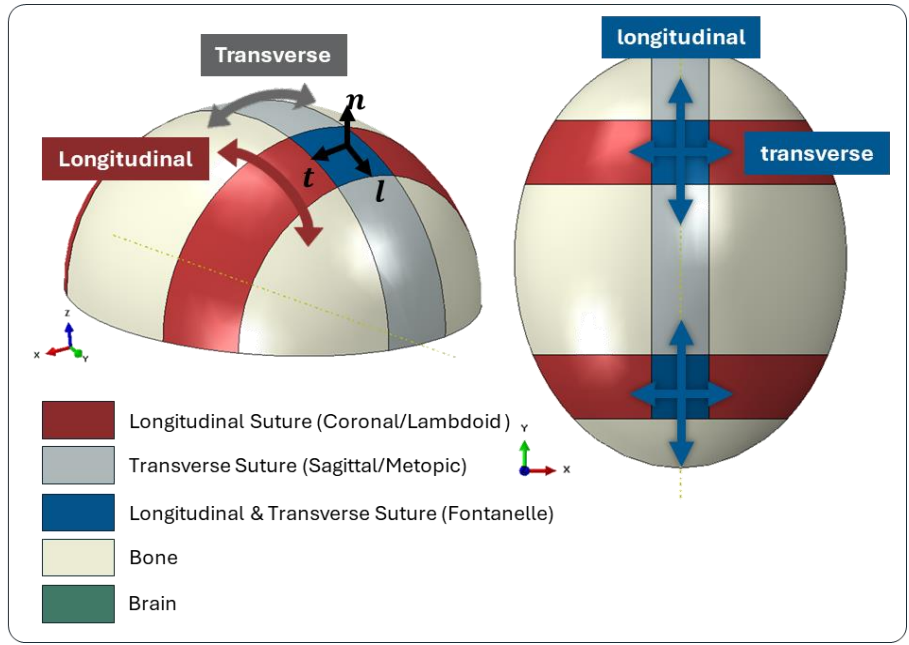

Figure 2 Local material orientation for skull sections and directional growth of sutures.

The governing equations for growth of the three tissues involved were described individually. ICV growth was implemented as a linear isotropic volumetric expansion that acts as the mechanical stimulus for growth initiation in bone and suture tissues. Stress relaxation takes place as a response to the growth happening in suture, and strain is shown to be decreasing during the growth process. Therefore, strain is an appropriate stimulus for growth law, as it aligns with the physical cellular level growth, which is activated by specific amount of stretch in cells [3,6]. Based on this theory, growth patterns for the suture and bone sections were described as a feedback-driven process. Here, the growth increment ($\Delta \boldsymbol{F}^g$) was driven by the deviation of the elastic deformation element ($\boldsymbol{F}^e$) in the growth direction, from a reference activation stretch ($\lambda$), scaled by a growth rate ($k$), and time step ($dT$) across the tissue local material directions ($i = l, t, n$). Mathematically, the growth increments for the three materials in the system were defined by the following sets of equations:

$$\Delta \boldsymbol{F}^g_{Brain} = k_{Brain}.dT \qquad (10)$$

$$\Delta \boldsymbol{F}^g_{Bone,i} = \left(\boldsymbol{F}^e_{Bone,i} - \lambda_{Bone}\right).k_{Bone,i}.dT, \qquad i = l, t, n \qquad (11)$$

$$\Delta \boldsymbol{F}^g_{Sut,i} = \left(\boldsymbol{F}^e_{Sut,i} - \lambda_{Sut}\right).k_{Sut,i}.dT, \qquad i = l, t, n \tag{12}$$

The brain's growth is modelled as a predetermined factor, resulting in a volume increase by a factor of two by the end of the first year [14,26]. Since the shape of the skull remains similar after growth in a healthy skull [6,18], the growth deformations in longitudinal and transverse directions for the suture tissue have a factor of $\frac{k_l}{k_t} = 1.5$.

Figure 3 shows the flowchart summarising the growth algorithm, implemented in ABAQUS (SIMULIA, Dassault Systémes) using UMAT subroutine. According to this algorithm, growth term is updated in each iteration, then the new elastic deformation is calculated, and based on the constitutive models of materials, Cauchy stresses are calculated and returned to the implicit solver for deformations and the updated configuration of the system.

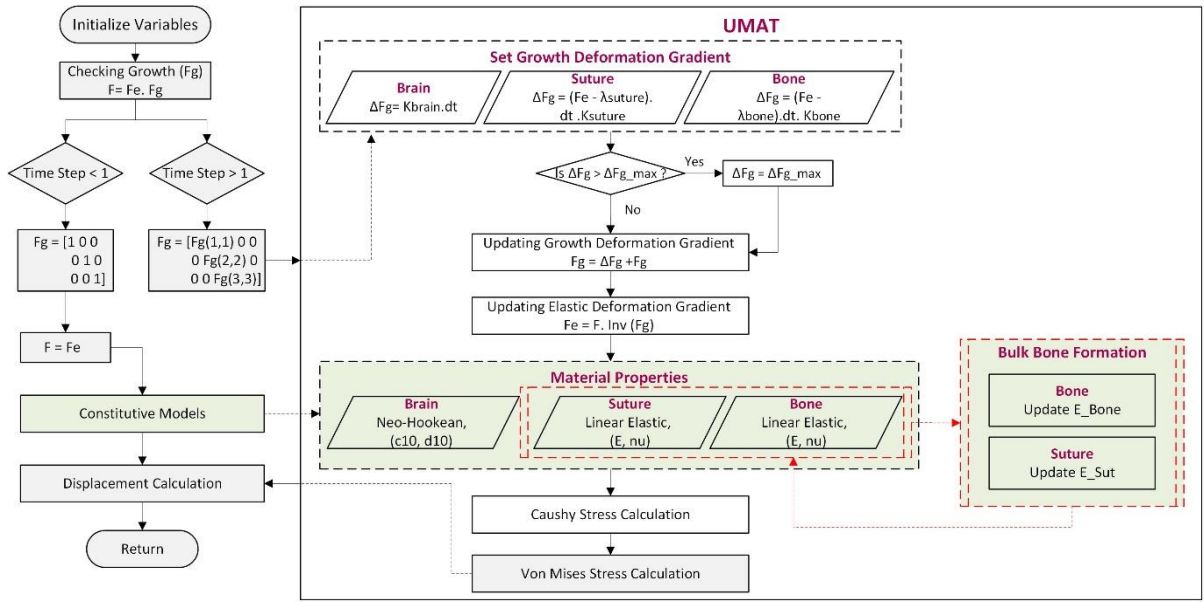

Figure 3 Flowchart of the growth algorithm applied on the skull assembly, as a UMAT.

## 2.4 Bone Formation

During growth, bone formation happens in the cranial vault from the bone edges that modifies the mechanical properties based on increases the elastic modulus of the suture. To represent the bone formation process within skull, bulk bone formation scheme [2,12] was applied on the whole suture elements. According to this scenario, the elastic modulus of the suture elements was increased monthly, at individual steps. Moreover, bone mechanical properties are variable during growth [27]. To accommodate this feature, the bulk mechanical properties of bone tissue also increased monthly in the algorithm. Therefore, the elastic modulus of the suture and bone occupied in the growth algorithm were updated over time using the following rule.

$$E_{updated} = (\lfloor T/n \rfloor + 1)\, E_{initial} \qquad (13)$$

In this equation, $T$ denotes growth time, $n$ shows the step time that $E$ needs to be updated (monthly here), and $E_{updated}$, and $E_{initial}$ represent the current and initial elastic modulus of the tissue, respectively.

### 2.5 Model Parameters

For the healthy skull geometry, the growth governing functions were implemented for individual material sections, with the growth parameters presented in Table 1.

Table 1 Growth model parameters

|  | Growth direction | Growth Rate ($k$) | Activation Stretch ($\lambda$) |
|---|---|---|---|
| ICV | $l, t, n$ | 0.0085 | n/a |
| Bone | $l, t$ | 2.0 | 1.0001 |
| Suture (lengthening) | $l$ | 3.75 ($1.5 k_{Sut,w}$) | 1.001 |
| Suture (transverse) | $t$ | 2.5 | 1.001 |

Isotropic, linear elastic materials were assumed for the bone and suture tissues. The initial material properties of the bone and suture were specified and updated during growth based on the bulk bone formation scenario, as described in [Section 2.4](). The elastic modulus of bone and suture tissues were increased by 250 MPa and 20 MPa in 10 intervals, respectively. These step values resulted in the final elastic modulus of 2,900 MPa and 230 MPa for bone and suture, respectively at the end of 12 month of growth [1,2]. The Poisson ratio was considered 0.22 and 0.3 for bone and suture, respectively [18]. The ICV tissue was considered as a hyperelastic Neo-Hookean material, with $C_{10}$ of 0.051 MPa, and $D_{10}$ factor of 0.0026 1/MPa. The material model parameters are summarised in Table 2.

Table 2 Material properties used for modelling [21].

| Tissue | $E_{initial}\,(MPa)$ | $\nu$ | $C_{10}(MPa)$ | $D_{10}\,(1/MPa)$ |
|---|---|---|---|---|
| Bone | 421 | 0.22 | N/A | N/A |
| Suture | 30 | 0.3 | N/A | N/A |
| ICV | N/A | N/A | 0.051 | 0.0026 |

### 2.6 Craniosynostosis Cases

With the developed growth model on a healthy skull, individual geometries of craniosynostosis cases were developed, to predict cranial morphology arising from these conditions. Figure 4 shows range of cases considered and how skull growth was restricted in specific sutures, representing suture fusion by the time of simulations. According to each synostosis condition, individual suture sections were inactivated, replaced by bone tissue properties, resulting in growth restrictions and therefore dysmorphology on the final shape of the developed skull. Dysmorphologies due to synostosis conditions were quantitively studied, by calculating the $CI$ for the deformed skulls, and normalising the elements' displacements on the synostosis cases, compared to the original shape.

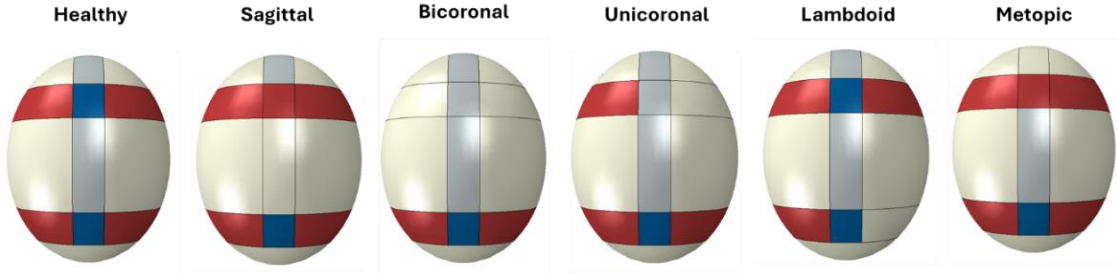

Figure 4 Healthy skull compared to the individual craniosynostosis cases, with inactivated (or fused) sutures.

### 2.7 Sensitivity Analysis

A set of sensitivity analyses was performed to assess the effect of the input parameters on the FE simulation results. The analysis consists of two main categories of independent parameters, including growth parameters and mechanical properties, applied to a healthy skull geometry, which is called the baseline model. Table 3 summarises the independent parameters and their values used for the sensitivity study. The mechanical properties were studied in terms of the effect of the elastic modulus of the bone and suture, and ICV tissue material properties. Moreover, the growth parameters were studied in terms of the growth rate in the suture, and bone tissues. The independent effects of the parameters were studied in terms of the changes on the growth pattern and the resulting geometry of the skull.

Table 3 Material and growth properties used in sensitivity analysis

|  | $E_{bone}$ $(MPa)$ | $E_{suture}$ $(Mpa)$ | $E_{brain}$ $(C_{10}, D_{10})$ | $k_{Sut}$ | $k_{Bone}$ |
|---|---|---|---|---|---|
| **Baseline model** | 421 | 30 | 0.3, 0.48 | 2.5 | 2 |
| **Bone stiffness** | 3000 | N/A | N/A | N/A | N/A |
| **Suture stiffness** | N/A | 100 | N/A | N/A | N/A |
| **Brain stiffness** | N/A | N/A | 0.3, 0.48 | N/A | N/A |
| **Suture growth** | N/A | N/A | N/A | 10 | N/A |
| **Bone growth** | N/A | N/A | N/A | N/A | 8 |

N/A indicates no change in the test model compared to the baseline model.

## 3 Results

### 3.1 Healthy Skull Growth

Figure 5(a) shows the growth curve of the total brain volume over time, which shows that the prescribed brain growth rate resulted in a two-fold volumetric growth with the rate of 100 cm³/month, over 12 months (form initial value of 1,260 cm³ to a final volume of 2,500 cm³). Figure 5(b) shows the developed healthy skull at the end of 12 months of growth, in which all the sutures are active and have grown in longitudinal and transverse directions. The new skull dimensions resulted in $CI$ of 77.73%, which is close to the initial $CI$, and in the range of a healthy skull shape [28]. Figure 5(c) shows how different sutures contributed to development of the skull shape after growth. The calculated growth deformation gradient terms show that longitudinal and transverse growth was almost homogeneous for a healthy case. Maximum transverse growth happened at the sagittal suture, as these elements experienced higher elastic deformation in this direction. Metopic suture transverse growth allowed for shaping the forehead in the model. These contours also show that suture growth was greater than growth in bone sections. Bone plates did not undergo large deformation or growth and therefore kept their original surface curvature after development.

Figure 5(d, e) show the Von Mises stress levels on the skull tissues and ICV section, respectively. Stress concentrations were observed on the bone edges connected to the sutures. These were in the order of 100 KPa, with stress concentration in the sagittal suture. Figure 5(f) show the predicted contact pressure experienced by the ICV-bone interface due to the skull constraint on the ICV. Intercranial pressure prediction showed higher pressures under longitudinal sutures, that indicated restricted growth in transverse directions in these regions. However, less pressure was experienced under Fontanella sutures, as they were associated with bi-directional growth.

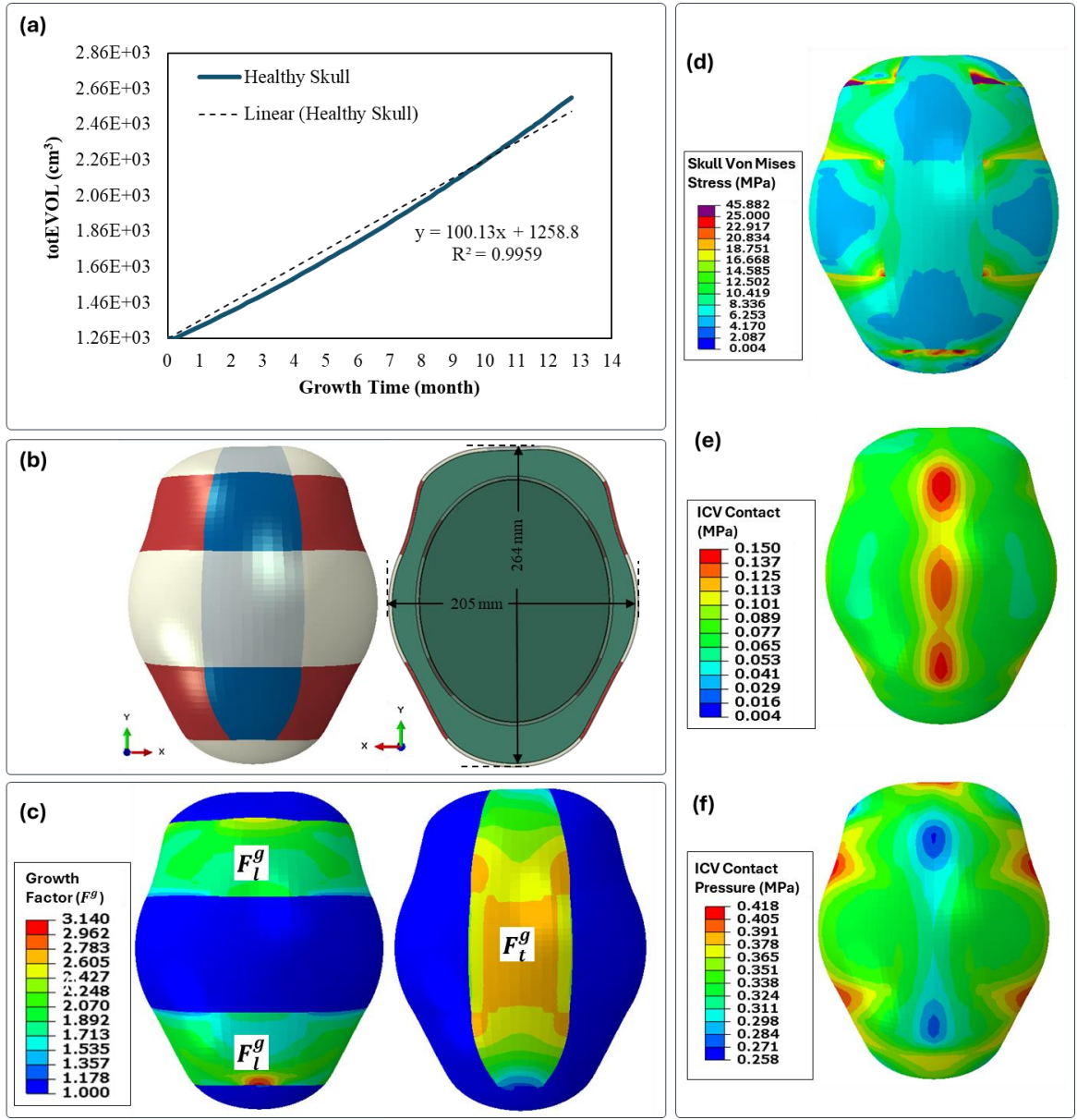

Figure 5 Results from a healthy skull growth model; (a) Skull growth rate in terms of volume increase per month, (b) Healthy skull after 12 months of growth and the new configuration and dimensions of skull, (c) growth deformation gradients in the two principal directions, longitudinal ($F_l^g$) and transverse ($F_t^g$) suture growths, (d) Von Mises stress distribution in MPa on the skull and (e) brain sections, (f) ICP in MPa on the brain tissue.

## 3.2 Craniosynostosis Cases

Figure 6 shows the craniosynostosis models and subsequent dysmorphologies that were predicted during the 12-month growth process. The normalised deformation of each synostosis case is also shown in Figure 6, which described how the skull has changed in comparison to its original shape. The maximum displacements were observed first on metopic and then sagittal synostosis cases, due to constraint growth in frontal region of the head and sagittal widening, respectively.

In the sagittal fusion case, significant lengthening of the skull was predicted, resulting in a narrow head shape. When both coronal sutures were closed in the bicoronal case, severe symmetric widening of the skull was predicted. In the unicoronal fusion, the open side suture bulged outward to accommodate the brain growth happening underneath. When one of the lambdoid sutures was fused, larger growth in the posterior fontanella and open sided lambdoid suture was observed, causing asymmetric deformity in posterior side of skull. Finally, in the metopic suture fusion, the forehead was developed with irregular shape as the bone plates were restricted and attached, causing brain to grow on the posterior side.

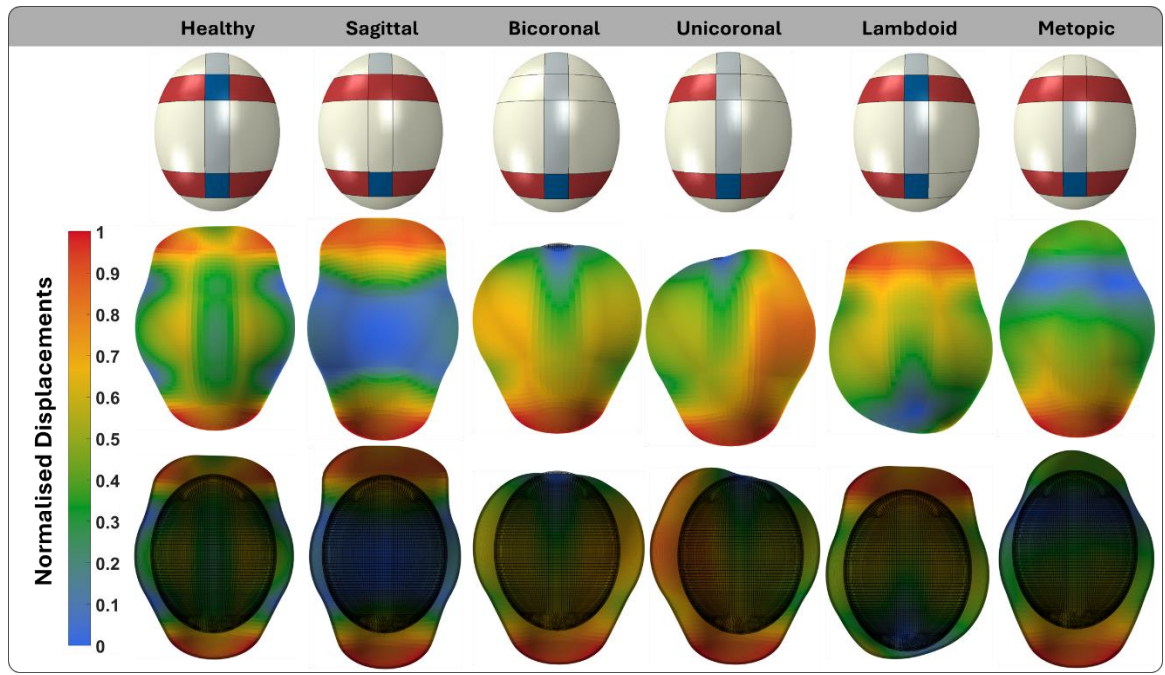

Figure 6 Dysmorphologies developed due to craniosynostosis cases individually. On top, there are FE models of craniosynostosis with corresponding fused sutures. Middle raw is the top view of normalized displacement contour, with respect to the maximum displacement in each simulation. The bottom raw is the bottom view of the normalised deformed configurations compared to the original skull shape.

Figure 7 shows the shape changes and growth deformation tensor elements over time for each case considered. According to the normalised displacement results presented in Figure 7(a), differences in skull shape started to appear from 6 months and became more apparent as growth continued. In the healthy case, displacements were homogenous over time across different directions, whereas in the metopic case, posterior side displacement magnitudes were higher compared to the anterior section elements.

Based on the growth deformation gradient results in Figure 7(b), in the healthy skull, growth in longitudinal directions were homogenous, with maximum growth developed at sagittal suture in transverse (widening) direction. In metopic-synostosis case, growth deformation gradients were developed non-homogenously to accommodate volumetric growth of the brain, in both longitudinal and transverse directions. In the transverse direction, maximum growth was observed on the posterior fontanelle suture, with the metopic suture fused. Comparing the two cases, longitudinal growth for the metopic case was higher compared to the healthy skull. In both cases, growth in bone tissue is negligible compared to the growth at sutures.

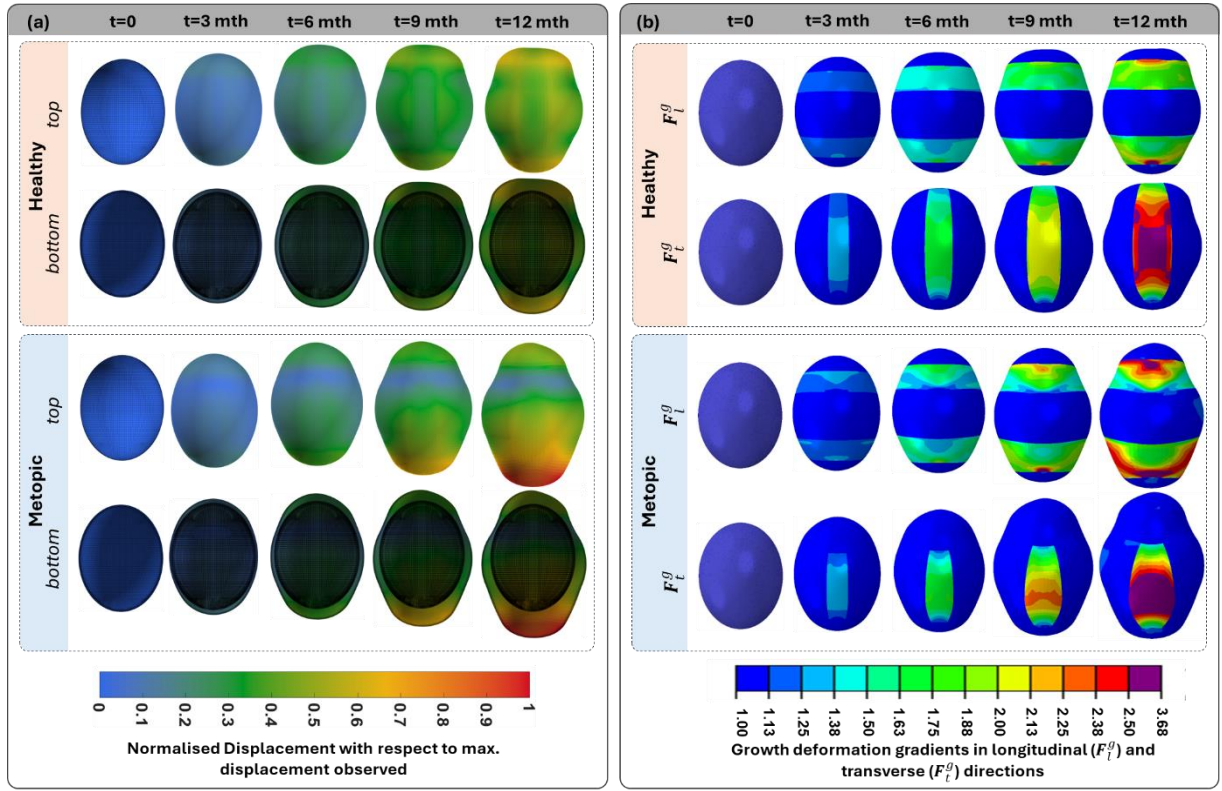

Figure 7 Growth results over time for the healthy and metopic-synostosis cases; (a) Normalized displacement over time during growth for healthy and metopic cases with respect to the maximum displacement observed in all simulations, from top and bottom views. (b) Growth deformation gradients over time in longitudinal ($F_l^g$) and transverse ($F_t^g$) directions, for healthy and metopic-synostosis cases.

Figure 8 shows the mechanical effect of growth on the skull and ICV, whereby Von Mises stress on the skull and the intercranial pressure experienced by ICV during growth are presented. According to stress distributions in Figure 8(a), the healthy skull has the minimum and most homogenous stress distribution, compared with each of the craniosynostosis cases, which confirms that under normal skull growth, suture growth releases the mechanical load on brain and skull. Maximum stress levels were recorded for metopic fusion, where the frontal (anterior) skull was restricted in transverse growth. In all the cases, stress concentrations were observed on the bone edge. Fontella suture areas showed lower stress concentrations in all models, as growth in both directions releases the stress in those regions.

Intercranial pressure predictions in Figure 8(b) shows that ICV area under fused suture experiences higher pressures for individual craniosynostosis cases. Compared to a healthy skull, as stress analysis also showed, the whole brain was exposed to higher pressure levels in craniosynostosis cases, with locally high pressurised regions. The maximum pressure on the brain was predicted in the metopic suture fusion case, in the frontal section of skull. Here, ICV had pressure values that were almost three times higher than in the health case.

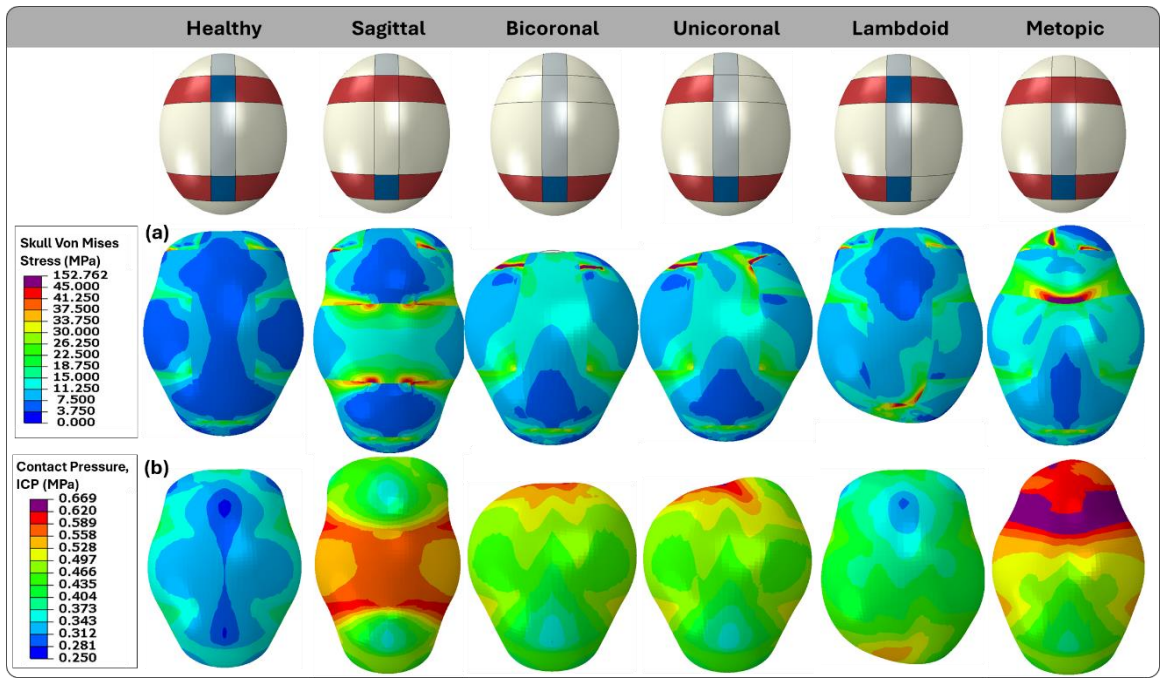

Figure 8 Results showing the mechanical environment on brain and skull; (a) Von Mises stress distribution on the healthy and craniosynostosis skulls, in MPa, (b) Comparison between intercranial contact pressure (ICP) across intercranial surface for all healthy and synostosis cases in MPa, after 12 months of growth.

### 3.3 Comparison with Clinical Data

Dysmorphologies due to individual craniosynostosis cases were compared with clinically observed dysmorphological skull shapes, as shown in Figure 9. $CI$ values were also quantitatively compared with the clinical $CI$ values, as shown in Table 4. According to the results, the dysmorphologies and models calculated $CI$ across all cases were in close agreement with clinical data. The algorithm predicted the overall shape of the dysmorphologies with maximum error of 12% for the metopic craniosynostosis case. These results show that the growth algorithm on the FE model is capable of predicting the overall malformed skull shape in individual craniosynostosis cases.

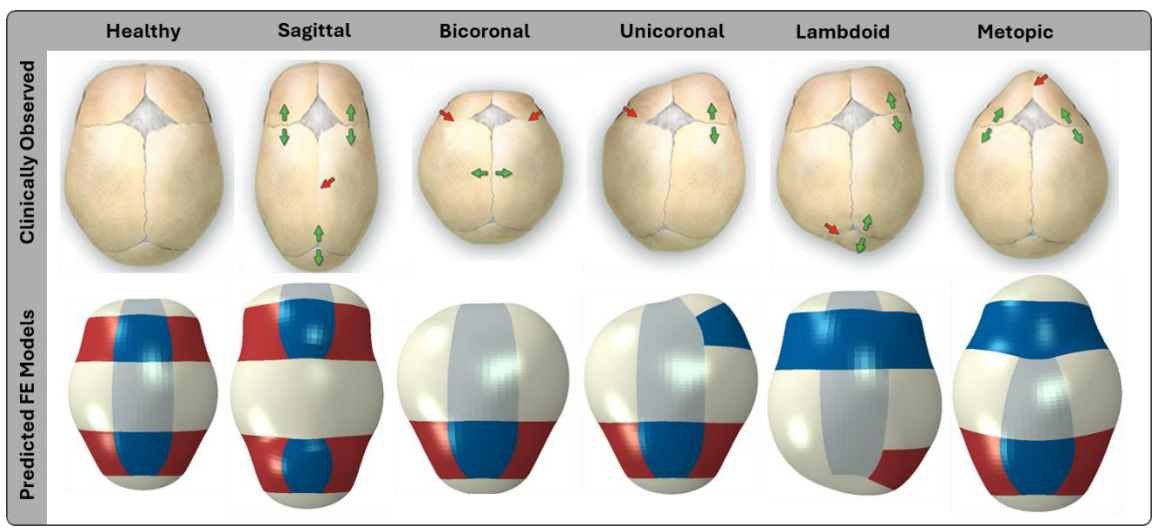

Figure 9 Comparison of clinically observed skull malformation due to craniosynostosis cases [recreated from [29]], with predicted dysmorphologies in this study via FE analysis.

Table 4 Dimensional measurements and cephalic index for the FE models, and clinical data [6].

|  | Healthy | Sagittal | Bicoronal | Unicoronal | Lambdoid | Metopic |
|---|---|---|---|---|---|---|
| **Length, l (mm)** | 264.01 | 290.58 | 246.10 | 256.19 | 245.15 | 282.27 |
| **Width, w (mm)** | 205.20 | 196.17 | 220.35 | 216.93 | 206.40 | 200.55 |
| **Model *CI*** | 77.73 | 67.51 | 89.54 | 84.68 | 84.19 | 71.05 |
| **Clinical *CI*** | 78.7 | 68.13 | 94.16 | 84.44 | 88.09 | 81.49 |

*3.4    Sensitivity Analysis*

The results of sensitivity study are presented in Figure 10, showing both the test cases and the baseline model, which serves as the control. The effects of parameters on the developed geometries are reported in Table 5, in terms of geometrical dimensions and $CI$.

*Increased bone stiffness:* For the case of stiffer bone tissue, the growth deformation gradients show lower growth in sutures in both directions as shown in Figure 10(a, b), and slightly higher in bone tissue. The resulting geometry of skull does not represent a normal skull shape, with slightly higher $CI$ compared to the normal case according to Table 5. Displacement contours in Figure 10(c) showed deformed parietal bone plates, with larger deformation magnitudes, widening the skull, and restricted growth in z-direction as shown in Figure 10(d), compared to the baseline model.

*Increased suture stiffness:* Increased suture stiffness restricted the elastic deformation in this tissue, however, resulted in larger growth deformation gradients as shown in Figure 10(a, b). The final shape of the skull was more elliptical with smooth curvatures in borders, compared to the baseline model. However, this parameter change did not affect the $CI$ according to Table 5.

*Increased brain stiffness*: Increased brain stiffness ($C_{10}$) resulted in larger growth deformation gradients on the suture sections, specifically on sagittal suture, as shown in Figure 10(a, b). Higher displacement values were predicted on the suture-bone elements on the edges in Figure 10(c). Also, due to higher stiffness of brain, there was separation between the skull and brain tissue, with the brain tissue not compliant enough to follow skull deformation and "fill in" the skull volume. Since the most substantial change in the geometry was shown on the out of plane direction ($z$-direction), the $CI$ does not reflect this effect, and results show minor changes compared to the baseline model, as reported in Table 5.

*Increased suture growth rate*: With increased growth rate in suture, larger growth deformation was captured, with the same elastic deformation term, as shown in Figure 10(a, b). Also, these results show larger displacement in the sagittal section, compared to the anterior and posterior regions on the skull. Faster growth in sutures cased a slight increase in the $CI$ measured, according to Table 5.

*Increased bone growth rate*: Increased growth rate in the bone section led to large growth deformations on bone plates, shown in Figure 10(a, b), such that bone plates had more displacement magnitudes, compared to the suture elements, as in Figure 10(c). The overall shape of the skull was also affected by the larger growth of bone sections, and sutures in circumferential region developed less growth, as shown in Figure 10(d). Larger growth of bone plates also affected the *CI* and overall geometry of the skull, as shown in Table 5.

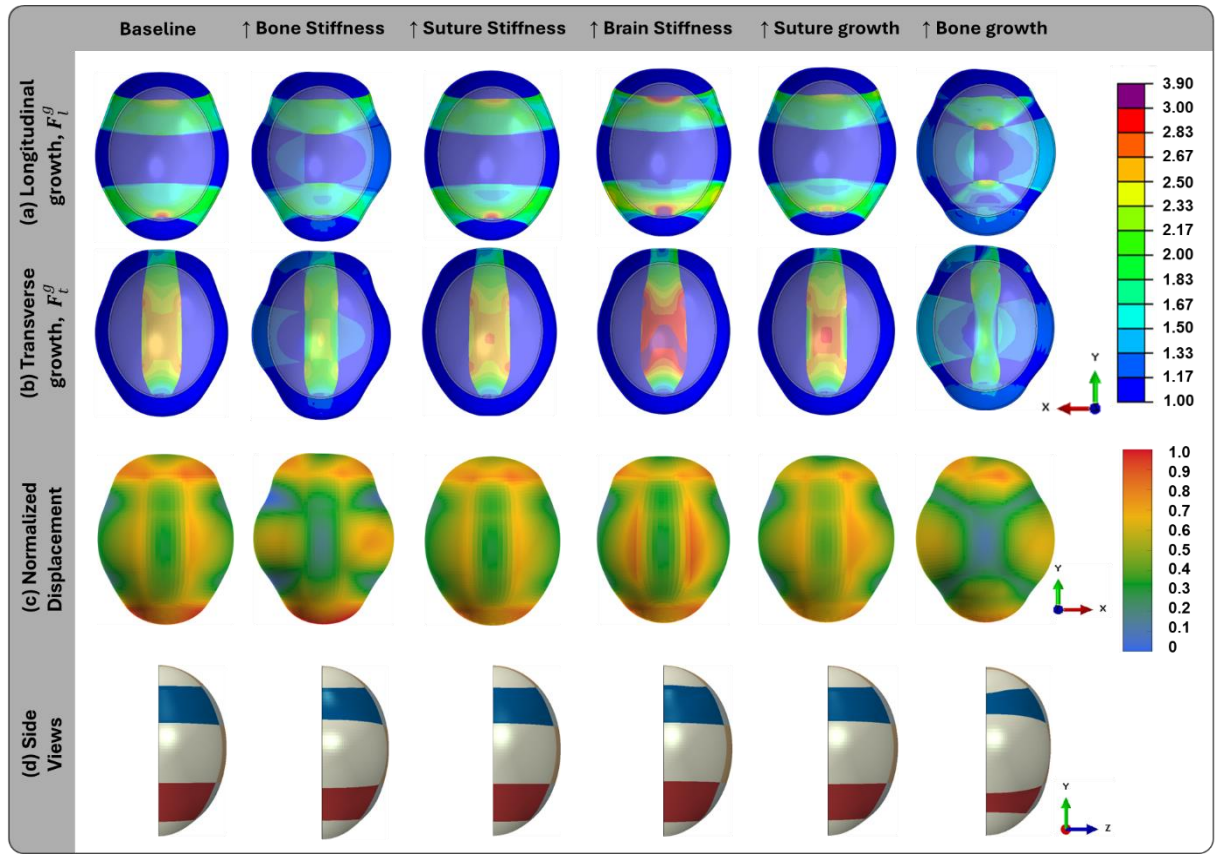

Figure 10 Sensitivity analysis results for mechanical properties and growth rate parameters; showing growth deformations in (a) longitudinal ($F_l^g$), (b) and transverse ($F_t^g$) directions, (c) normalized displacement for each test, on a healthy skull, and (d) side view of the skull sections after 12-months growth.

Table 5 Dimensional measurements and cephalic index for the FE models in sensitivity analysis

|  | **Baseline model** | **Bone stiffness** | **Suture stiffness** | **Brain stiffness** | **Suture growth** | **Bone growth** |
|---|---|---|---|---|---|---|
| **Length, l (mm)** | 257.9 | 259.2 | 257.2 | 252.7 | 252.8 | 255.8 |
| **Width, w (mm)** | 204.0 | 212.4 | 203.5 | 200.6 | 205.5 | 212.0 |
| **Model CI** | 79.1 | 82.0 | 79.1 | 79.4 | 81.3 | 82.9 |

## 4 Discussion

This study developed a physically-based, mechanically driven growth model that captures the dynamic interplay between brain expansion, skull deformation, and bone formation over the first year of life, both in healthy development and in craniosynostosis cases. A central contribution of this model is its coupled and continuous simulation of ICV and skull growth, driven by volumetric expansion of the brain. Unlike models that apply static or uniform pressures [4,15], this approach allows tissue-level feedback through a strain-based algorithm, better mimicking the physiological mechanisms by which brain expansion drives sutural bone growth and demonstrating how the growing brain interacts with the skull reconstructions [2,10]. Importantly, the model demonstrated that a mechanically regulated growth mechanism, without the inclusion of explicit biochemical signals, can sufficiently predict realistic skull geometries. This supports the hypothesis that mechanical signalling plays a pivotal role in cranial development, particularly in regulating sutural growth and guiding the formation of calvarial shape in response to brain expansion [3,5,6]. The model provided excellent predictions of resulting dysmorphologies in craniosynostosis cases, showing good agreement with clinical observations for known phenotypes [29], highlighting the model's utility in evaluating pathological skull development and informing surgical planning.

The simulation results presented here demonstrate that the physically-based growth model can accurately reproduce the compensatory growth mechanisms and characteristic dysmorphologies associated with various forms of craniosynostosis. Unlike previous studies on craniosynostosis that have largely focused on bone formation mechanism [7,12,21], our finite element (FE)-based growth framework is able to predict skull formation in the first 12 months of life, driven by biological and mechanical growth cues. Previous computational studies have typically relied on static evaluation growth approaches to describe skull shape changes [13,15–18]. While these approaches have been useful for craniosynostosis correction assessments, they lack predictive capacity and long-term biomechanical assessments, particularly in simulating how premature suture fusion leads to compensatory deformation over time. Typically, these approaches use simplified volumetric expansion [4,19] to describe general deformation trends but fail to capture the complex, anisotropic growth behaviour across the sutures and skull base. The results of our FE-based growth simulations shows that the model can predict dysmorphologies due to premature suture fusion in craniosynostosis conditions, which closely match clinically observed dysmorphologies [29]. For example, in cases where longitudinal suture growth is fused, transverse growth compensates to accommodate ICV expansion as seen in bicoronal craniosynostosis, while the opposite occurs in sagittal craniosynostosis. Comparison of the model 3D skull shape prediction this computational model with clinical observations in Figure 9, confirms the model and algorithm accuracy in capturing craniosynostosis-related developmental disorders. These cases demonstrate how craniosynostosis disrupts normal cranial expansion patterns, with the model accurately predicting that when growth in one direction is constrained, the skull adapts by expanding orthogonally to maintain ICV expansion. Specific dysmorphologies, such as excessive longitudinal growth in sagittal case, transverse growth in bicoronal case, forehead thinning in metopic case, and asymmetric bulging in unicoronal and lambdoid cases, align with individual dysmorphologies in clinical data, and are reflected in the model's *CI* as well. Some discrepancies between simulation and clinical data were noted. For instance, the fontanella suture were bulged outward towards *z*-direction in several synostosis cases, that was not recorded in clinical data. This difference can be explained by discretised growth deformation elements on different suture sections in the models, which results in discontinuous growth on the adjacent tissues. In this regard, more precise quantitative comparison can be made using other indices, such as cranial H/B, H/L and interocular index [26], that involves out of plane measurements of the skull (z-direction in this model), or cranial module (CM), or craniofacial index (CI) that work better on a CT-based cranial model [14].

Although there are other approaches to predict dysmorphologies due to craniosynostosis conditions, such as data-driven models [30], few studies were able to present a physiological based growth pattern that occurs [6]. In this study, we extended modelling capabilities beyond the mechanically driven growth algorithm by incorporating variable mechanical properties that evolve over time and by developing a coupled growth model that simultaneously accounts for ICV tissue and skull reconstruction, rather than relying solely on applied pressure for simplifying the mechanical environment from brain [4,6]. The growth model in this study simulates gradual bone growth over 12 months in a single step. Other similar studies applied discrete growth over multiple steps [1,8,20], primarily to control volume change. Depending on the step types, loading conditions on the skull reconstructions due to brain growth might not be consistence throughout the total growth time in these models. However, our study considered a continuous linear growth of brain in a single step, throughout the simulation, that mimics the slow and continuous biological growth. The developed model in this study enables a long-term assessment of the system, whereas in other computational models, only short-term growth periods were studied [10]. According to the mechanical environment results in Figure 5, stress concentration on the bone-suture interface comes from the stiffness difference of the two tissues, also, due to the lower growth deformations on bone plates compared to the suture sections. Coronal and lambdoid sutures, which are active in longitudinal growth only, on the edges in connection with bone plates were shown to be under large stress values (Figure 5(d)), which could be explained by their restricted growth in transverse direction. This stress could have been released by transverse growth, that is now trying to keep the bone plates in place from widening (transverse) effect of the sagittal suture. Also, the higher stress concentration on bone plates edges, can be interpreted as physiological mechanical stimuli for the bone formation mechanism, that is captured by the model. The stress contours provide valuable insights for optimizing the osteotomy location in correction surgery. While contact separation was reported in other similar studies [9], the model developed in this study maintained continuous contact between all skull elements and the brain surface, with no open gaps observed during the growth process. Also, the model measured intracranial pressure (ICP) values at approximately 2 kPa, which, when compared to the literature KPa [15], it overestimates brain pressure by two orders of magnitude. This can be explained by simplifications applied on the model in terms of material properties, and merging material levels into one ICV section in the model.

The model's reliance on input parameters was assessed through a sensitivity analysis, evaluating each parameter's impact relative to a baseline model. The analysis showed negligible effects on the cephalic indices (CI), indicating that parameter variations had minimal impact on the final geometry. Changes in material properties altered growth patterns, as illustrated in Figure 10, due to the strain-feedback mechanism in the growth algorithm. For instance, increasing bone stiffness enhanced growth by elevating stress and strain stimuli. A stiffer ICV material resulted in isotropic deformation, unaffected by skull or suture constraints, while a softer ICV material conformed to surrounding bone deformation. The study highlighted that overly stiff ICV material prevents filling of the skull chamber, detaching from adjacent tissues, whereas sufficiently soft ICV material follows bone deformation. Balanced growth rates among the three interacting tissues were critical, as imbalances led to dysmorphologies or over-constrains and excessive tissue stress, as shown in Figure 10. Growth in bone plates was minimal compared to sutures (Figure 7), yet it facilitated stress relaxation and improved simulation convergence. The model's sensitivity to boundary conditions, such as $x$- and $y$-fixed constraints, influenced skull curvature, particularly in bicoronal and metopic cases, necessitating consideration of their impact on displacements.

This research has certain limitations, outlined as follows. First, the material constitutive models employed in this study did not account for the viscoelastic properties of the cranial bones, suture and brain tissues. These properties could improve the mechanical adaptability of the system during growth [31], and relaxes stresses induced from growth deformations on the tissues. Moreover, this study did not incorporate the bone remodelling process on the bone surface within the model. This process is responsible for the changes in bone plates curvatures, change in the thickness of skull and contributes to the final shape of skull [6,32]. Furthermore, the model did not include localized bone formation processes to simulate suture fusion. Instead, a bulk bone formation approach was utilized, which accounted for temporal changes in the mechanical properties of bone and suture. Nevertheless, this bulk bone formation method has been shown to effectively predict the overall skull morphology [12]. Lastly, the study employed a simplified elliptical skull geometry. A model more closely resembling the physiological skull shape, incorporating dura mater and additional bone plates, could better represent the system. These elements were excluded from this research. Nonetheless, the geometric simplification did not compromise the study's primary focus on developing a mechanically induced growth algorithm. As the results demonstrate, the model successfully captured key geometrical indices. This model provides the necessary tools to further study and optimise the correction techniques, time, size and location of the craniotomy for an optimal outcome of surgical interventions in craniosynostosis treatment.

## 5  Conclusion

This study presented a physically-based computational model of a coupled skull growth and volumetric ICV growth with a strain-feedback loop growth governing tissue deformation in sutures and bones, capturing mechanically driven growth dynamics. The algorithm was also used to predict skull malformation due to early fusion of sutures in individual craniosynostosis cases and showed quantitative and qualitative agreement to clinical data. The coupled growth of brain and skull reconstructions during development in the model allows assessment of their mechanical interactions, and mechanical consequences in each craniosynostosis cases. The model is potential to predict useful data towards more accurate planning for the surgical intervention in terms of location, time of intervention, as well as correction techniques.

# 6   Data availability

The original contributions presented in the study are included in the article. Further inquiries can be directed to the corresponding author.

# 8 Funding

This project has received funding from the European Union's Horizon Europe research and innovation programme under grant agreement No 101047008 (BIOMET4D). Views and opinions expressed are however those of the author(s) only and do not necessarily reflect those of the European Union or the European Innovation Council and SMEs Executive Agency (EISMEA). Neither the European Union nor the EISMEA can be held responsible for them.